\documentclass[12pt]{article}
\usepackage{graphicx}
\usepackage{amsmath}
\usepackage{amsfonts}
\usepackage{amssymb}
\usepackage{graphicx}
\newcommand{\be}{\begin{equation}}
\newcommand{\ee}{\end{equation}}
\newcommand{\bea}{\begin{eqnarray}}
\newcommand{\eea}{\end{eqnarray}}
\newcommand{\ba}{\begin{array}}
\newcommand{\ea}{\end{array}}

\newcommand{\cL}{{\cal L}}

\newcommand{\cA}{{\cal A}}
\newcommand{\cB}{{\cal B}}
\newcommand{\cM}{{\cal M}}
\newcommand{\cT}{{\cal T}}

\newcommand{\no}{\nonumber}
\newcommand{\lsim}{\stackrel{<}{_\sim}}

\newcommand{\ket}[1]{\vert {#1} \rangle}
\newcommand{\bra}[1]{\langle {#1}}
\newcommand{\la}{\langle}
\newcommand{\ra}{\rangle}

\def\plb#1#2#3{    {Phys. Lett.}~B {\bf #1}, #3 (#2)}

\def\prd#1#2#3{    {Phys. Rev.}~D {\bf #1}, #3 (#2)}

\def\prl#1#2#3{    {Phys. Rev. Lett. }{\bf #1}, #3 (#2)}
\def\ptp#1#2#3{    {Prog. Theor. Phys. }{\bf #1}, #3 (#2)}

\def\ijmpa#1#2#3{  {Int. J. Mod. Phys. }~A {\bf  #1}, #3 (#2)}

\def\dfrac#1#2{{\displaystyle \frac{#1}{#2}}}

\def\slash#1{\setbox0=\hbox{$#1$}\dimen0=\wd0                    
      \setbox1=\hbox{/} \dimen1=\wd1 \ifdim\dimen0>\dimen1
      \rlap{\hbox to \dimen0{\hfil/\hfil}} #1                        
      \else                                       
      \rlap{\hbox to \dimen1{\hfil$#1$\hfil}}  
      /   \fi}                                         
\def\simge{\mathrel{\rlap{\raise 0.511ex \hbox{$>$}}{\lower 0.511ex 
\hbox{$\sim$}}}}
\def\simle{\mathrel{\rlap{\raise 0.511ex \hbox{$<$}}{\lower 0.511ex 
\hbox{$\sim$}}}} 
\def\slash#1{\setbox0=\hbox{$#1$}\dimen0=\wd0                    
      \setbox1=\hbox{/} \dimen1=\wd1 \ifdim\dimen0>\dimen1
      \rlap{\hbox to \dimen0{\hfil/\hfil}} #1                  
      \else                                       
      \rlap{\hbox to \dimen1{\hfil$#1$\hfil}}
      /   \fi}

\newcommand{\nn}{\nonumber}

\def\b{\begin{equation}}
\def\e{\end{equation}}

\newcommand{\beq}{ \begin{equation}}
\newcommand{\eeq}{ \end{equation}}
\newcommand{\beqau}{\begin{eqnarray*}}
\newcommand{\eeqau}{\end{eqnarray*}}
\newcommand{\beqa}{\begin{eqnarray}}
\newcommand{\eeqa}{\end{eqnarray}}

\newcommand{\Lam}{\Lambda_{\rm QCD}}

\newcommand{\as}{\alpha_{ s}}

\begin{document}

\begin{flushright}
February 2005 \\
RM3-TH/05-2\\
hep-ph/0503107 \\
\end{flushright}
\thispagestyle{empty}
\setcounter{page}{0}

\vskip   2 true cm 

\begin{center}
{\Large \textbf{Light-quark Loops in $K \to \pi \nu\bar\nu$}} \\ [20 pt]
\textsc{Gino Isidori},${}^{1}$ \textsc{Federico Mescia},${}^{1,2}$ 
and \textsc{Christopher Smith},${}^{1}$  \\ [20 pt]
${}^{1}~$\textsl{INFN, Laboratori Nazionali di Frascati, I-00044 Frascati,
      Italy} \\ [5 pt]
${}^{2}~$\textsl{Dip. di Fisica, Universit\`a di Roma Tre, 
     Via della Vasca Navale 84, I-00146 Rome, Italy  } 

\vskip   2 true cm 

\textbf{Abstract}
\end{center}
\noindent
We present a comprehensive analysis of the contributions 
to $K \to \pi \nu\bar\nu$ decays not described by the leading 
dimension-six effective Hamiltonian. These include both
dimension-eight four-fermion operators generated at the charm 
scale, and genuine long-distance contributions which 
can be described within the framework of chiral perturbation 
theory. We show that a consistent treatment of the latter 
contributions, which turn out to be the dominant effect,
requires the introduction of new chiral operators already 
at $O(G_F^2 p^2)$. Using this new chiral Lagrangian, we analyze 
the long-distance structure of $K \to \pi \nu\bar \nu $ amplitudes 
at the one-loop level, and discuss the role of the dimension-eight
operators in the matching between short- and long-distance components.
From the numerical point of view, we find that these  $O(G_F^2 \Lambda^2_{\rm QCD})$ 
corrections enhance the SM prediction of $\cB(K^+\to\pi^+\nu\bar\nu)$
by about $\approx 6\%$.
\setcounter{footnote}{0}

\newpage

\section{Introduction}
Flavour-changing neutral-current (FCNC) processes and the related 
GIM mechanism \cite{GIM} are one of the most fascinating aspects 
of flavour physics. Within the Standard Model (SM), FCNC amplitudes are 
strongly suppressed and often completely dominated by short-distance 
dynamics. In this case, their precise study allows 
to perform very stringent tests of the model and ensures 
a large sensitivity to possible new degrees of freedom \cite{My,Buras:2004}.

On general grounds, we can distinguish two types of FCNC 
transitions: those where the leading short-distance amplitude
exhibits a power-like GIM mechanism, and those 
where the GIM suppression is only logarithmic.
This distinction plays a key role in kaon physics, where 
long-distance effects are enhanced by the hierarchy
of the Cabibbo-Kobayashi-Maskawa (CKM) matrix \cite{CKM}
and could easily spoil the short-distance structure 
of FCNC amplitudes. Only in the case of a power-like GIM mechanism, 
or a power-like suppression of light-quark contributions, 
long-distance effects can be kept under good theoretical control.
   
The quark-level transition $s \to d \nu \bar \nu$ 
is the prototype of FCNC amplitudes with a power-like GIM mechanism.
The leading contributions to this amplitude are genuine
one-loop electroweak effects, which are usually encoded 
in the following  $O(G_F^2)$ effective Hamiltonian 
(see e.g. Ref.~\cite{Buras:2004}):
\be
{\cal H}^{(6)}_{eff}=\frac{G_F}{{\sqrt 2}}\frac{\alpha}{ 2\pi \sin^2\theta_W} \left[ 
 V^{\ast}_{ts}V_{td} X_t (x_t) +  V^{\ast}_{cs}V_{cd} X^l_c (x_c) \right]
 (\bar s d)_{V-A}(\bar\nu_l \nu_l)_{V-A}~,
\label{eq:heff} 
\ee
with $x_q = m_q^2/M_W^2$
As a consequence of the power-like GIM mechanism, the coefficient 
functions of the unique dimension-6 operator in Eq.~(\ref{eq:heff}) behave 
as $X_q (x_q) \propto  x_q$ (up to logarithmic and subleading 
power corrections).\footnote{~This behavior illustrates the 
$O(G_F^2)$ structure of ${\cal H}^{(6)}_{eff}$:
$G_F \alpha/( 2 \sqrt 2 \pi \sin^2\theta_W)\times x_q = G^2_F m_q^2 /(2\pi^2)$}
This implies that the term proportional to $V^{\ast}_{ts}V_{td}$, 
which is enhanced by the large top-quark mass and can be precisely 
computed in perturbation theory at the electroweak scale \cite{BB_old,MU,BB_new}, 
is the dominant contribution.

In the case of CP-conserving transitions, such as the $K^+\to\pi^+\nu\bar\nu$ 
decay, the charm contribution 
in ${\cal H}^{(6)}_{eff}$ cannot be neglected. Indeed the power suppression 
of $X^l_c$ with respect to $X_t$ is partially compensated 
by the large CKM coefficient ($|V^{\ast}_{cs}V_{cd}| \approx 10^3 \times 
|V^{\ast}_{ts}V_{td}|$). Moreover, charm quarks remain dynamical 
degrees of freedom for a large range of energies below the 
electroweak scale. This imply an enhancement factor due to 
large logs and a stronger sensitivity to QCD corrections in $X^l_c$. 
Thanks to the  NLO calculation of Ref.~\cite{BB_old,BB_new}, 
$X^l_c$ is known with a relative precision of about $18\%$.
Since the charm contribution amounts to about $30\%$
of the total magnitude of $\cA(s\to d\nu\bar\nu)_{\rm SM}$, 
the NLO uncertainty translates into an error 
of about $10\%$ in the SM estimate of $\cB(K^+\to\pi^+\nu\bar\nu)$. 
This type of uncertainty can possibly be reduced to below $4\%$ 
with a NNLO calculation of $X^l_c$ \cite{Buras:2004}. 
 
Aiming to get a few \% precision on  the $K^+\to\pi^+\nu\bar\nu$ 
amplitude, it becomes important to address the question 
of the subleading terms not described by the effective 
Hamiltonian in Eq.~(\ref{eq:heff}). In particular, 
${\cal H}^{(6)}_{eff}$ does not allow to evaluate 
in a systematic way the contributions of 
$O(G_F^2  \Lam^2)$ 
to the $K^+\to\pi^+\nu\bar\nu$ amplitude.
Naively, these  contributions are parametrically
suppressed only by $O(\Lam^2/m_c^2) \approx O(10\%)$ with respect 
to the charm contribution in Eq.~(\ref{eq:heff}).
Therefore, they are not trivially negligible
at the few \% level of accuracy. 
The purpose of this paper is a systematic analysis
of this type of effects. 

\medskip

The relevant subleading contributions to $K\to\pi\nu\bar\nu$ 
can  be safely  computed in the limit $V_{td}=0$ (or in the limit 
where charm- and up-quark loops appear with the same CKM 
coefficient) and can be divided into two groups:
\begin{enumerate} 
\item[i.] $O(1)$ (tree-level) matrix elements
of local FCNC operators of dimension eight, such as 
$(\bar s\Gamma \partial s) \times (\bar \nu \Gamma \partial \nu)$,
appearing in the $O(G^2_F)$ Hamiltonian;
\item[ii.]  $O(G_F)$ (beyond tree-level) matrix elements of the 
 $\Delta S=1$ dimension-6 four-fermion operators appearing 
in the  $O(G_F)$  effective Hamiltonian.
\end{enumerate} 
The distinction between these two types of effects depends 
on the choice of the renormalization scale for the effective 
four-fermion theory. 
For instance, choosing a renormalization scale well above the charm mass, 
one can essentially neglect the dimension-8 FCNC operators 
and encode all the effects via appropriate matrix-elements of the 
$\Delta S=1$ effective Hamiltonian (where charm quarks 
are still treated as dynamical degrees of freedom). This approach 
would be the most natural choice in view of a calculation 
of these matrix elements by means of lattice QCD (probably the 
ultimate way to  address this problem). Waiting 
for such a calculation on the lattice, here we adopt 
a different procedure and choose a renormalization scale for the 
effective four-fermion operators below the charm mass.
As shown in Ref.~\cite{BI,Falk}, this is the most natural choice 
in view of a fully analytic approach to the problem. 

Concerning the construction of the dimension-8 four-fermion 
Hamiltonian, we completely confirm the results of Ref.~\cite{Falk}.
However, we substantially extend this work by analysing the 
impact of the genuine long-distance component of the amplitude,
namely the matrix-elements of $\Delta S=1$ four-fermion operators 
(where only $u$, $d$ and $s$ quarks are treated as dynamical 
degrees of freedom). The latter component cannot be computed 
at the partonic level and the best analytic approach to evaluate 
its size is provided by chiral perturbation theory (CHPT). 
Several authors have already addressed the issue of long-distance   
effects in the $K^+\to \pi^+\nu\bar\nu$ amplitude within 
the framework of CHPT \cite{RS,HL,Lu,Geng,Fajfer}. 
However, as we shall show, all previous attempts to address 
this problem were not complete and, in particular, were not 
able to discuss the matching between short-  and long-distance 
components of the amplitudes.

The paper is organized as follows: in section~\ref{sect:Heff}
we analyse the structure of the dimension-8 four-fermion Hamiltonian. 
The main new results are contained in section~\ref{sect:CHPT},
where we construct the effective Lagrangian relevant to 
evaluate FCNCs of $O(G_F^2)$, we evaluate the
long-distance components of $K\to \pi\nu\bar\nu$ amplitudes 
up to $O(G_F^2p^4)$, and we discuss the matching between 
the chiral approach and the four-fermion operators.
These results are used in section~\ref{sect:Num} to 
address the numerical impact on $\cB(K^+\to \pi^+\nu\bar\nu)$.
The conclusion are summarized in 
section~\ref{sect:concl}.

\section{The $O(G_F^2)$ four-fermion effective Hamiltonian}
\label{sect:Heff}

\begin{figure}[t]
\begin{center}
\includegraphics[width=9cm,,height=5.5cm]{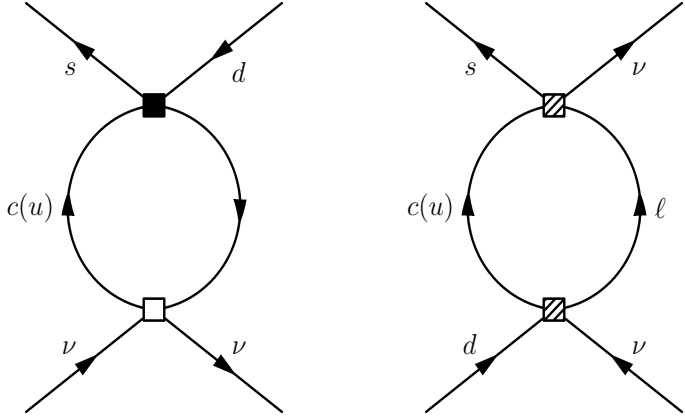}
\end{center}
\caption{\it One-loop diagrams corresponding to the 
T-products in Eqs.~(\ref{o1N})--(\ref{obN}).}
\label{fig:fullmc}
\end{figure}

Since we are interested only in contributions generated 
by up- and charm-quark loops (namely we neglect the corrections 
of $O(\Lambda_{\rm QCD}/m_t^2)$), we can set $V_{td}=0$.
In this limit, CKM unitarity allows to express all the 
relevant contributions in terms of one independent 
CKM combination: $\lambda_c =  V^{\ast}_{cs}V_{cd} = - V^{\ast}_{us}V_{ud}$.
As discussed in Ref.~\cite{BB_old,Falk}, 
the central point for the construction of the 
low-energy effective theory is the expansion 
in terms of local operators of the following T-products,\footnote{~For a complete discussion, 
we refer to Ref.~\cite{BB_old}. Note that, since we are interested
also in the subleading terms arising by the expansion of the T-products,
we include both left-handed and vector components of
$Q_Z^{qq \nu\nu}$ in Eq.~(\ref{eq:QZ}). The latter 
has been ignored in \cite{BB_old} since it does not 
contribute to the leading dimension six operator.}
\bea
\label{o1N} O^Z_1 &=& -i\int d^4x\ T \left[
        Q_1^{cc}(x)\ Q_Z^{cc\nu\nu}(0)\,- 
        Q_1^{uu}(x)\ Q_Z^{uu\,\nu\nu}(0)\right]~,  \\
\label{o2N} O^Z_2 &=&
   -i\int d^4x\ T \left[
        Q_2^{cc}(x)\ Q_Z^{cc\,\nu\nu}(0)\,- 
        Q_2^{uu}(x)\ Q_Z^{uu\,\nu\nu}(0)\right]~,  \\
\label{obN} O_l^B &=&
   -i\int d^4x\ T \left[ 
       Q^{cl}(x)\ Q^{lc}(0)\ -
       Q^{ul}(x)\ Q^{lu}(0)\right]~,
\eea
whose leading term is given by
\be
Q_l^{(6)} =  \bar s \gamma^\mu (1-\gamma_5) d ~ \bar\nu_l \gamma_\mu (1-\gamma_5) \nu_l~.
\ee
Here  
\bea
Q_1^{qq} &=& \bar s_i \gamma^\mu (1-\gamma_5) q_j ~ \bar q_j \gamma_\mu (1-\gamma_5) d_i~, \nn  \\
Q_2^{qq} &=& \bar s_i \gamma^\mu (1-\gamma_5) q_i ~ \bar q_j \gamma_\mu (1-\gamma_5) d_j~, 
\label{eq:Q12}
\eea
denote the leading $\Delta S=1$ four-quark operators ($q=u,c$), 
\be 
Q_Z^{qq \nu\nu} = \bar q_k \gamma^\mu \left[ (1-\gamma_5)
-\frac{8}{3} \sin^2\theta_W  \right]
 q_k ~ \bar\nu_l \gamma_\mu (1-\gamma_5) \nu_l \label{eq:QZ}
\ee 
is the effective neutral-current coupling induced by the integration of the $Z$ boson, and 
\bea
Q_3^{ql} &=& \bar s \gamma^\mu (1-\gamma_5) q  ~ \bar \nu_l
                     \gamma_\mu (1-\gamma_5) l \nn \\
Q_4^{lq} &=& \bar l \gamma^\mu (1-\gamma_5) \nu_l ~ \bar q \gamma_\mu (1-\gamma_5) d 
\label{eq:Qbox}
\eea
are the effective charged-current couplings induced by integration 
of the $W^\pm$ bosons. Note that, even if we are interested in 
dimension-8 operators, we work at $O(G_F^2)$ and we can safely use 
a point-like propagator in the case of both $Z$ and $W^\pm$
bosons. The T-products in Eqs.~(\ref{o1N})--(\ref{obN}) 
correspond to the diagrams in Figure~\ref{fig:fullmc}.

The first two steps necessary for the construction of the effective theory, 
namely the determination of the initial conditions at $\mu=M_W$  
of $O^Z_{1,2}$, $O_l^B$ and $Q^{(6)}$, and the renormalization group evolution 
down to lower scales, proceeds exactly as in Refs.~\cite{BB_old}-\cite{BB_new}.
On the other hand, we differ from these works in the last step, namely 
the removal of the charm as dynamical degrees of freedom. 
In this case we proceed as in Ref.~\cite{Falk}, matching the 
operator product expansion of the T-products into an 
effective theory which includes also dimension-8 operators. 
The structure of the local terms, for $\mu_{IR} \lsim m_c$,
takes the form of the following effective Hamiltonian density 
\be
{\cal H}^{(6+8)}_{eff}(\mu_{IR})=\frac{G_F}{{\sqrt 2}}\frac{\alpha}{ 2\pi \sin^2\theta_W}
\lambda_c \sum_{l=e,\mu,\tau} \left[ X^l_c (x_c) Q^{(6)}_l + \frac{1}{M_W^2} \sum_{i} 
C^l_i(\mu_{IR})  Q^{(8)}_{il}  \right]~.
\label{eq:heff_68} 
\ee
Neglecting neutrino masses, the only $Q^{(8)}_{il}$ with non-vanishing 
coefficients to lowest order in $\alpha_s(m_c)$ are
\bea
\label{o81} Q_{1l}^{(8)} &=&  \bar s \gamma^\mu (1-\gamma_5) d ~ 
\partial^2 \left[\bar\nu_l \gamma_\mu (1-\gamma_5) \nu_l\right]~, \no \\
\label{o82} Q_{2l}^{(8)} &=&  (\bar s {\overleftarrow D}_\alpha) \gamma^\mu (1-\gamma_5) 
({\overrightarrow D}_\alpha d) ~ 
\bar\nu_l \gamma_\mu (1-\gamma_5) \nu_l~, \no \\
\label{o83} Q_{3l}^{(8)} &=&  (\bar s {\overleftarrow D}_\alpha) \gamma^\mu (1-\gamma_5) d ~ 
\left[ \bar\nu_l ({\overleftarrow{\partial^\alpha}} 
- {\overrightarrow{\partial^\alpha}} )\gamma_\mu(1-\gamma_5)\nu_l \right]~.
\label{eq:Q1l}
\eea
The operator $Q_{1l}^{(8)}$ arises by the neutral-current coupling 
(left diagram in Figure~\ref{fig:fullmc}), while $Q_{2l}^{(8)}$ and  
$Q_{3l}^{(8)}$ are generated by the charged-current coupling 
(right diagram in Figure~\ref{fig:fullmc}). The operator
$Q_{3l}^{(8)}$, which has been considered first in Ref.~\cite{BI}, 
is the only term which can induce a CP-conserving contribution 
to the $K_2 \to \pi^0 \nu_l \bar \nu_l$  transition. 
In agreement with the results of Ref.~\cite{BI,Falk}, we find 
\bea
 C_{1}^l(\mu_{IR}) &=&
 \frac{1}{12}\,\left(1-\frac{4}{3}\sin^2\theta_W\right)
 \log\left( m_c^2/\mu_{IR}^2\right)\left[3 C_{1}(\mu_c) +C_{2}(\mu_c)\right] 
 \nn\\
 C_{2}^{e,\mu}(\mu_{IR}) \, &=&   \frac{1}{2}
 \log\left( m_c^2/\mu_{IR}^2\right)\, C_B(\mu_c) \label{eq:Cdim8}\\
 C_{2}^\tau(\mu_{IR}) \, &=&  -\frac{1}{4}
 f\left( m_c^2/m_{\tau}^2\right)\, C_B(\mu_c)\nn \\
 C_{3}^{l}(\mu_{IR}) \, &=& \,- C_{2}^{l}(\mu_{IR}) \nn
\eea
where $C_{1,2}(\mu_c)$ represent the Wilson coefficients
at scale $\mu_c={\cal O}(m_c)$ for the operators 
in Eqs.~(\ref{o1N})--(\ref{o2N}), and 
\be
f(x)=\left(\frac{6x-2}{(x-1)^3}-2\right)\log x -\frac{4x}{(x-1)^2}\,.
\ee
In the calculation of the subleading dimension-8 operators, 
we shall take into account QCD corrections only up to the 
leading logarithmic level. In this approximation, 
the $C_{1,2,B}(\mu_c)$ coefficients reads
\bea
\!\!\!\!\!\!\!\!\!\!\!\!\!\!\!\! &&
C_{1}(\mu_c) = \frac{1}{2}\left[\left(\frac{\as(m_c)}{\as(m_b)}\right)^{-6/25}
   \left(\frac{\as(m_b)}{\as(M_W)}\right)^{-6/23}-\left(\frac{\as(m_c)}{\as(m_b)}\right)^{12/25}
   \left(\frac{\as(m_b)}{\as(M_W)}\right)^{12/23} \right] \nn\\
\!\!\!\!\!\!\!\!\!\!\!\!\!\!\!\! && 
C_{2}(\mu_c) = \frac{1}{2}\left[\left(\frac{\as(m_c)}{\as(m_b)}\right)^{-6/25}
   \left(\frac{\as(m_b)}{\as(M_W)}\right)^{-6/23}+\left(\frac{\as(m_c)}{\as(m_b)}\right)^{12/25}
   \left(\frac{\as(m_b)}{\as(M_W)}\right)^{12/23} \right] \nn\\
\!\!\!\!\!\!\!\!\!\!\!\!\!\!\!\!
&& C_{B}(\mu_c) = 1 
\eea
Before leaving this section, we comment on the role of the scale $\mu_{IR}$.
This scale works as an infrared cutoff for the expansion of the T-products 
in Eqs.~(\ref{o1N})--(\ref{obN}):  $\mu_{IR}$  separates the
long-distance contributions associated to up-quark loops (with low virtuality), 
from the local part encoded in ${\cal H}^{(6+8)}_{eff}(\mu_{IR})$.
In order to cancel the $\mu_{IR}$ dependence in the physical amplitudes, 
we should sum to $\bra{\pi\nu\bar\nu}|{\cal H}^{(6+8)}_{eff}(\mu_{IR})\ket{K}$
also the non-local contribution generated by the matrix 
elements of the five four-fermion operators in Eqs.~(\ref{eq:Q12})--(\ref{eq:Qbox}), 
with $q=u$. In these matrix elements $\mu_{IR}$ should act as ultraviolet cut-off
for the light degrees of freedom. The estimate of these matrix 
elements will be addressed in section~\ref{sect:CHPT}.

\subsection{Matrix elements of the dimension-8 operators}
The contributions of the dimension-8 operators 
can be conveniently normalized in terms of the leading 
matrix element of $Q_{l}^{(6)}$. The simplest case 
is the one of $Q_{1l}^{(8)}$, for which we can write 
 \be
 \la \pi^+(k)\nu_l \bar\nu_l \vert Q_{1l}^{(8)} \vert K^+(p) \ra = 
  - q^2 ~ \la \pi^+\nu_l \bar\nu_l \vert Q_{l}^{(6)} \vert K^+ \ra
 \ee
where $q = (p-k)^2 =(p_\nu + p_{\bar\nu})^2$. 

Concerning $Q_{2l}^{(8)}$ and  $Q_{3l}^{(8)}$, 
we can proceed as in Ref.~\cite{BI} finding a suitable 
chiral representation for the corresponding 
bilinear quark currents. The contribution of $Q_{3l}^{(8)}$, 
which describes the transition into a $\ket{\nu\bar\nu}$ final state 
with $J=2$, turns out to be completely negligible~\cite{BI}.
This matrix element i)~suffers of a severe kinematical suppression;
ii)~vanishes to lowest order in the chiral expansion;
iii)~does not interfere with the leading term. 
On the contrary, $Q_{2l}^{(8)}$ generates a non-negligible
contribution; however, this cannot be expressed in terms of 
known low-energy couplings. In general, we can write 
 \be
 \la \pi^+(k)\nu_l \bar\nu_l \vert Q_{2l}^{(8)} \vert K^+(p) \ra = 
{\hat B}_{2}\left[ p\cdot k + O(m_q) \right] ~ \la \pi^+\nu_l \bar\nu_l \vert Q_{l}^{(6)} \vert K^+ \ra
\label{eq:hatB2}
 \ee
where ${\hat B}_{2}$ is an unknown hadronic parameter, 
expected to be of $O(1)$, and $ O(m_q)$ denotes contributions 
proportional to light-quark masses.

In summary, the leading contributions to the 
$K^+\to\pi^+\nu\bar\nu$ amplitude generated by 
the effective Hamiltonian in (\ref{eq:heff_68}) 
can be written as 
\be
\cA^{(8)} ( K^+ \to \pi^+ \nu_l \bar\nu_l )
= - \la \pi^+\nu_l \bar\nu_l \vert {\cal H}^{(6+8)}_{eff}(\mu_{IR}) \vert K^+ \ra 
\equiv \cA_Z^{(6)} + \cA_Z^{(8)} + \cA_{WW}^{(8)} 
\ee
where,  adopting the standard CHPT convention 
$\la \pi^+ \vert \bar s \gamma^\mu d \vert K^+ \ra = - (p + k)^\mu$
and using $\lambda = -\lambda_c$, we have 
\bea
 \cA^{(6)} & = &  - \frac{G_F}{{\sqrt 2}}\frac{\alpha~ \lambda }{ 2\pi \sin^2\theta_W} 
 X^l_c (x_c) ~\left[ (p+k)^\mu ~\nu_l \gamma_\mu (1-\gamma_5) \nu_l \right] \label{eq:Pc_d} \\
 \cA_Z^{(8)} & = &  \phantom{+} \frac{G_F}{{\sqrt 2}}\frac{\alpha~ \lambda }{ 2\pi \sin^2\theta_W} 
\frac{q^2}{M_W^2}  C_1^l(\mu_{IR}) ~\left[ (p+k)^\mu ~\nu_l \gamma_\mu (1-\gamma_5) \nu_l \right] \\
 \cA_{WW}^{(8)} & = & - \frac{G_F}{{\sqrt 2}}\frac{\alpha~ \lambda }{ 2\pi \sin^2\theta_W} 
  {\hat B}_{2} \frac{p\cdot k}{M_W^2}
  C_2^l(\mu_{IR}) ~\left[ (p+k)^\mu ~\nu_l \gamma_\mu (1-\gamma_5) \nu_l \right]
\label{eq:A8summ}
\eea

\section{$K\to \pi\nu\bar\nu$ amplitudes within CHPT}
\label{sect:CHPT}

As discussed in the previous section, the scale dependence 
of  $K\to \pi\nu\bar\nu$ amplitudes 
induced by the dimension-8 operators must be compensated by 
a corresponding scale 
dependence of their long-distance component.
The latter is generated by the matrix elements of four-fermion 
operators which involve only light quarks ($u$,$d$ and $s$)
and light lepton fields. 
In this case both internal and external fields do not 
involve high-energy scales, thus a partonic calculation 
of this part of the amplitude would be inadequate. 
In the following we shall present an estimate 
of these contributions in the framework of CHPT.

The four-fermion operators we are interested in are 
four-quark operators of the type in Eq.~(\ref{eq:Q12}) as well 
as quark-lepton couplings of the type in Eqs.~(\ref{eq:QZ})--(\ref{eq:Qbox}). 
All these effective operators are generated by the exchange 
of a single heavy gauge boson ($Z$ or $W$), and correspondingly have an
effective coupling of $O(G_F)$. However, our final goal is the evaluation 
of their T-product between $\ket{K}$ and $\ket{\pi\nu\bar\nu}$
states which --by construction-- is of $O(G^2_F)$. 
As we shall show in the next subsection, this observation has important 
consequences in the framework of CHPT. In particular, 
it implies that a consistent treatment of these effects 
requires the introduction of new appropriate chiral operators 
of $O(G^2_Fp^2)$.

\subsection{The $O(p^2)$ chiral Lagrangian 
including $O(G^2_F)$ FCNCs}

We start by the considering the chiral realization of the 
$O(G_F)$ coupling between quark and lepton currents.  
To this purpose, we introduce the so-called strong chiral 
Lagrangian of $O(p^2)$ in presence of external 
currents~\cite{GL}\footnote{~The symbol $\langle\rangle$ denotes 
the trace over the $3\times3$ flavour space.}
\be
\cL_S^{(2)} = \frac{F^{2}}{4} \left\langle D_{\mu}U D^{\mu}U^{\dagger}\right\rangle 
+\frac{F^{2}B}{2}\left\langle \cM U+U^{\dagger} \cM \right\rangle~.
\label{eq:LagrQCD}
\ee
As usual, we define 
\be
U =\exp(\sqrt{2} i\Phi/F)~, \qquad \Phi=\left[
\begin{array}{ccc}
\frac{\pi^0}{\sqrt{2}}+\frac{\eta}{\sqrt{6}} & \pi^+ & K^+ \\
\pi^- & -\frac{\pi^0}{\sqrt{2}}+\frac{\eta}{\sqrt{6}} & K^0 \\
K^- & \bar K^0 & -\frac{2\eta}{\sqrt{6}} \\
\end{array}\right]
\ee
where $\cM={\rm diag}(m_u, m_d, m_s)$ and, to lowest order, 
we can identify $F$ with the pion decay constant
($F\approx 92\, {\rm MeV}$) and express $B$ 
in term of meson masses [$m^2_{\pi}=B(m_u+m_d)$]. 
The generic covariant derivative, $D_\mu U = \partial_\mu U -i r_\mu U +i U l_\mu$,
allows to systematically include the coupling to 
external currents ($l_\mu$ and $r_\mu$) transforming 
as $(8_L,1_R)$ and $(1_L,8_R)$ under $SU(3)_R \times SU(3)_L$. 
In the specific case of the $O(G_F)$ couplings to charged 
lepton currents, we can thus identify the covariant derivative with 
\be
D^{(W)}_{\mu}U  = \partial_{\mu}U  - i \frac{g}{\sqrt{2}}  U ( T_+ W^+_\mu + \mbox{\rm h.c.})~, 
\qquad 
T_+=\left(\ba{ccc} 0 & V_{ud} & V_{us} \\ 0 & 0 & 0 \\ 0 & 0 & 0 \ea \right)
\label{Wfield}   
\ee
The situation is slightly more complicated in the case of the $Z$ boson, 
which couples to both right- and left-handed $SU(3)$ currents, 
but also to the following singlet current
\be
 J^{(1)}_{L\mu} = \sum_{q=u,d,s}  \bar q_L \gamma_\mu q_L 
\ee
which has a non-vanishing $U(1)_L$ charge. 
As pointed out in \cite{Lu}, the $Z$ coupling to $J^{(1)}_{L\mu}$ involve 
a new effective coupling which cannot be determined 
within the $SU(3)_R \times SU(3)_L$ chiral group. 
Putting all the ingredients together, the covariant 
derivative with {\em external} $W$ and $Z$ fields reads 
\be
D^{(W,Z)}_{\mu}U  = \partial_{\mu}U - ig_{Z} Z_{\mu}\left(  \sin^{2}\theta_{W} \left[Q,U\right]  + UQ
 - \frac{a_{1}}{6} U \right) - i \frac{g}{\sqrt{2}}  U ( T_+ W^+_\mu + \mbox{\rm h.c.})~, 
\label{eq:Dmu}
\ee
where $Q= {\rm diag}(2/3,-1/3,-1/3)$, $g_{Z}=g/\cos\theta_{W}$ and $a_{1}$ 
is the new coupling related to the $U(1)_L$ current.\footnote{~The 
normalization of the $U(1)_L$ coupling is such that $a_{1}\to 1$ 
in the limit where we can extend the symmetry of the QCD action 
(with 3 massless quarks) from  $SU(3)_R \times SU(3)_L$ to 
$U(3)_R \times U(3)_L$, as it happens for $N_c\to \infty$.} 
We finally recall that, being external fields, the $W^\mu$ and 
$Z^\mu$ operators appearing in Eq.~(\ref{eq:Dmu}) are only a short-hand 
notation to denote the corresponding leptonic currents (as obtained 
after the integration of the heavy gauge boson). In particular,
the case we are interested in corresponds to 
\be
W^+_\mu  \to \frac{ g}{2 \sqrt{2} M_W^2 } \sum_l~\bar{l} \gamma_\mu (1-\gamma_5) \nu_l~, \qquad
  Z_\mu \to  \frac{ g_Z}{4 M_Z^2 } \sum_l \bar{\nu}_l \gamma_\mu (1-\gamma_5) \nu_l~. \label{eq:Zeff}
\ee

We shall now proceed by considering the $O(G_F)$ four-quark operators.
The chiral realization of the $|\Delta S=1|$ non-leptonic Hamiltonian 
has been widely discussed in the literature (see e.g.~Ref.~\cite{nonlept}).
The contraction of a single $W$ field coupled to quark currents leads to two independent 
structures  which transform 
as $(8_L,1_R)$ and  $(27_L,1_R)$, respectively. Given the strong phenomenological 
suppression of the $(27_L,1_R)$ terms, in the following we 
shall consider only the $(8_L,1_R)$ operators. 
To lowest order in the chiral expansion, $O(p^2)$, the 
$(8_L,1_R)$ non-leptonic weak Lagrangian contains only one term:
\be
\cL^{(2)}_{ \left|  \Delta S \right| =1 }  = 
G_{8} F^{4} \left\langle \lambda_{6} D^{\mu}U^{\dagger} D_{\mu}U\right\rangle
\label{eq:LW2}
\ee
Here $G_8=O(G_F)$ is the effective coupling which is usually fixed 
from $K\to2\pi$ and incorporates the phenomenological $\Delta I=1/2$ enhancement
($G_8 \approx 9\times 10^{-6}~{\rm GeV}^{-2}$), while 
$(\lambda_{6})_{ij}=\delta_{2i}\delta_{j3}+\delta_{3i}\delta_{j2}$.
As shown in Ref.~\cite{KMW,EKW}, the number of independent operators 
increase substantially at $O(p^4)$.
 
The inclusion in the non-leptonic chiral Lagrangian of 
external $SU(3)_R \times SU(3)_L$ currents --which are hidden 
in the covariant derivative in Eq.~(\ref{eq:LW2})--
allows to describe in a systematic way also non-leptonic 
weak interactions in presence of external gauge fields. 
This is for instance the case of the $K\to \pi \gamma$ 
amplitude analysed in Ref.~\cite{EPR}.  However, 
contrary to what stated in Ref.~\cite{Lu},
this minimal coupling is not sufficient 
in the $K\to\pi\nu\bar\nu$ case. 
Here we are interested in FCNC amplitudes
where the leptonic current is associated to a broken 
generator of the electroweak gauge group: at $O(G_F^2)$
this coupling is not anymore protected by gauge invariance.
This argument can easily be understood by looking 
at the effective Hamiltonian in Eq.~(\ref{eq:heff}): 
from the point of view of chiral symmetry, the FCNC dimension-6 
operator in Eq.~(\ref{eq:heff}) corresponds to a
non-gauge-invariant coupling between a  $(8_L,1_R)$ 
quark current and an external neutral left-handed current. 
We thus need to extend the basis of chiral operators  
and include the chiral realization of all the 
$O(G_F^2)$ independent terms of this type. 
At $O(p^2)$ the situation is again quite simple 
since we have only two independent terms:
\be
\left\langle \lambda_{6} U^{\dagger} D^{\mu} U l_\mu \right\rangle~,  \qquad  \qquad 
\left\langle \lambda_{6} U^{\dagger} D^{\mu} U \right\rangle  \left\langle l_\mu \right\rangle~.
\label{eq:newops}
\ee
In the specific case we are interested in, 
we can thus add to the $|\Delta S|=1$ Lagrangian 
in Eq.~(\ref{eq:LW2}), with the covariant derivative 
(\ref{eq:Dmu}), the following $O(G_F^2 p^2)$ term 
\be
\cL^{(2)}_{\rm FCNC}  = i  g_Z F^4 Z_{\mu}\, \left[ G^Z_8 \left\langle \lambda_{6} 
U^{\dagger} D^{\mu} U Q \right\rangle  
+ G^Z_1 a_1 \left\langle \lambda_{6} U^{\dagger} D^{\mu} U  \right\rangle 
\right]~,
\label{eq:LW2_FCNC}
\ee
where again the $Z^\mu$ field has to be understood 
as the neutral current in Eq.~(\ref{eq:Zeff}). 
 
The consistent inclusion of all the relevant $O(G_F^2p^2)$ operators 
has forced us to introduce two new effective chiral couplings of $O(G_F)$, 
namely $G^Z_8$ and $G^Z_1$. At this order the chiral Lagrangian 
$\cL^{(2)}_{\rm FCNC}+\cL^{(2)}_{ \left|  \Delta S \right|=1 }$
--with the covariant derivative given in (\ref{eq:Dmu})--
has a local current-current structure identical to the one of the leading 
dimension-6 Hamiltonian in Eq.~(\ref{eq:heff}). Since we are 
able to compute explicitly the $K^0\to Z$ matrix element 
in both approaches, we can fix the unknown chiral couplings 
by an appropriate matching condition on the $K^0\to Z$ amplitude.
In particular, imposing the condition
\be
\langle K^0 | \cL^{(2)}_{ \left|  \Delta S\right| = 1  } + \cL^{(2)}_{\rm FCNC} | Z^\mu \rangle =0
\label{eq:match}
\ee
we eliminate from the $O(G_F^2p^2)$ chiral Lagrangian any contamination 
from the leading dimension-6 operators. The matrix elements of the latter must then be computed 
directly by means of the partonic Hamiltonian, as in the standard approach. 
The condition (\ref{eq:match}) implies $G^Z_8 = -2 G_8$ and $G^Z_1 = G_8/3$,
which leads to the following structure for the 
weak Lagrangian of $O(G_F p^2 + G^2_F p^2)$:
\be
\cL^{(2)}_{ \left|  \Delta S \right| =1  +  {\rm GIM} } = 
G_{8} F^{4} \langle \lambda_{6}  \left[  D^{\mu}U^{\dagger} D_{\mu}U 
- 2 i g_Z Z_{\mu} U^{\dagger} D^{\mu} U \left( Q - \frac{a_1}{6}\right) \right] \rangle 
\label{eq:LW2_tot}
\ee
Using this Lagrangian,\footnote{~We denote 
it with a subscript ``GIM'' since the 
condition (\ref{eq:match}) is equivalent to enforcing an
exact GIM cancellation for the tree-level FCNC
associated to the leading short-distance operator.} 
together with the ordinary  $\cL_S^{(2)}$,
we are finally ready to analyse the structure of $K\to \pi\nu\bar\nu$
long-distance amplitudes up to the one-loop level.

\subsection{$Z$-mediated amplitudes: general structure and $O(p^2)$ results}
Following the same conventions adopted for the short-distance calculation, 
we define the $K(p) \to \pi(k) \nu \bar\nu$ amplitude mediated by $Z$ exchange by 
\be
\cA( K \to \pi \nu \bar\nu )_Z = -i \la \pi \vert 
~\frac{\delta}{\delta Z_\mu} \int [d\phi] e^{i \int dx \cL[\phi,Z] } ~ \vert K \ra  \times 
\frac{ g_Z}{4 M_Z^2 } \sum_l \bar{\nu}_l \gamma_\mu (1-\gamma_5) \nu_l
\ee
and we find it convenient to decompose it as follows
\be
\cA( K^i \to \pi^i \nu \bar\nu )_Z = \frac{G_F}{\sqrt{2}} G_8 F^2 
\left[ \cM_L^i~p^\mu + \cM_V^i \left(q^2 p^\mu -  p\cdot q~q^\mu \right)\right] 
\sum_l \bar{\nu}_l \gamma_\mu (1-\gamma_5) \nu_l
\label{eq:dec_M}
\ee
where $q=p-k$ and the form factors $\cM_{V,L}^i$  are regular functions 
in the limit $q^2\to 0$ and $p\cdot q\to 0$.
The $\cM_V^i$ terms, which can be different from zero only 
at $O(p^4)$ in the chiral expansion, correspond to the conserved
component of the coupling to the external current.

\begin{figure}[t]
\begin{center}
\includegraphics[width=11 cm]{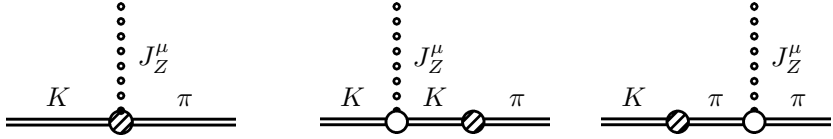}
\end{center}
\caption{\it Tree-level contributions to the $K\to \pi Z$ amplitudes          
within CHPT. Dashed and empty circles correspond to vertices derived 
from $\cL^{(2)}_{ \left|  \Delta S \right| =1  +  {\rm GIM} }$
and $\cL_S^{(2)}$, respectively}
\label{fig:tree}
\end{figure}

\begin{figure}[t]
\begin{center}
\includegraphics[width=15 cm]{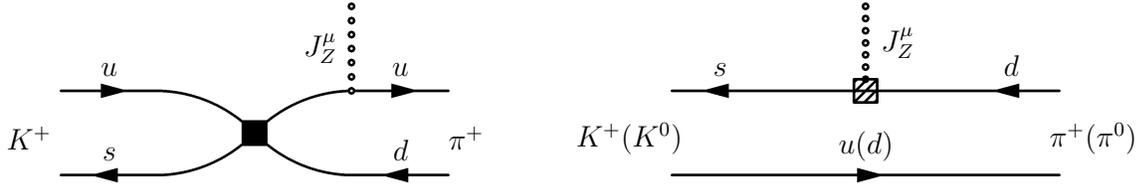}
\end{center}
\caption{\it Examples of valence-quark topologies appearing in 
 $K \to \pi  Z$ amplitudes. Left: non-spectator topology
contributing to the $K^+ \to \pi^+ $ transition  only
(the black box denotes a weak four-quark operator). 
Right: penguin-type contraction or insertion of the 
local FCNC effective coupling.}
\label{fig:spect}
\end{figure}

In the decomposition (\ref{eq:dec_M}) we have implicitly neglected 
the $O(q^\mu)$ terms which do not appear in the coefficient of $ \cM_V^i$
(all the $O(q^\mu)$ terms give a negligible result when 
contracted with the neutrino current). This allows to substantially 
simplify the calculation and, in particular, to neglect the 
bilinear couplings $Z^{\mu}\partial_{\mu}\phi$. In this limit, the only 
non-vanishing  $O(p^2)$ tree-level diagrams contributing  
the $K \to \pi Z$ amplitude are those in figure~\ref{fig:tree}.
They leads to the following results for the non-conserved terms
of charged and neutral channels:
\be
\cM_L^{+(2)} = 4~,  \qquad  \cM_L^{0(2)} = 0~.
\label{eq:tree}
\ee 
Because of the modified weak Lagrangian in (\ref{eq:LW2_tot}),
these results are qualitatively different from those available 
in the literature. On the one side, find that the charged amplitude ($K^+ \to \pi^+ Z$) 
is different from zero and is completely determined in terms of known couplings.
In particular, the amplitude does not vanish in the large $N_C$ limit, 
as claimed in Ref.~\cite{Lu}. On the other hand, we find that the 
neutral amplitude ($K^0 \to \pi^0 Z$) is identically zero.
By comparison, it should be noted that the $K^0 \to \pi^0 Z$ amplitude 
computed with the minimal-coupling prescription of Ref.~\cite{Lu} 
is different from zero, even in the large $N_C$ limit.
The vanishing of the $K^0 \to \pi^0 Z$ amplitude at $O(p^2)$ 
is a direct consequence of the condition 
imposed by Eq.~(\ref{eq:match}), which removes from the low-energy 
effective Lagrangian the spurious tree-level FCNC coupling 
generated by the minimal substitution. As already 
mentioned, by means of this procedure the leading 
penguin-type contractions (see figure~\ref{fig:spect}) 
are removed from the low-energy effective theory and treated 
directly by means of the partonic Hamiltonian.
In this approach, the only non-vanishing contributions of $O(p^2)$ 
are the genuine long-distance effects associated to 
non-spectator topologies,
which can affect only the charged transition $K^+ \to \pi^+ Z$
 (see figure~\ref{fig:spect}).

\subsection{One-loop contributions to $K \to \pi  Z$ amplitudes}

\begin{figure}[t]
\begin{center}
a)\hskip 1 cm  \includegraphics[width=10 cm]{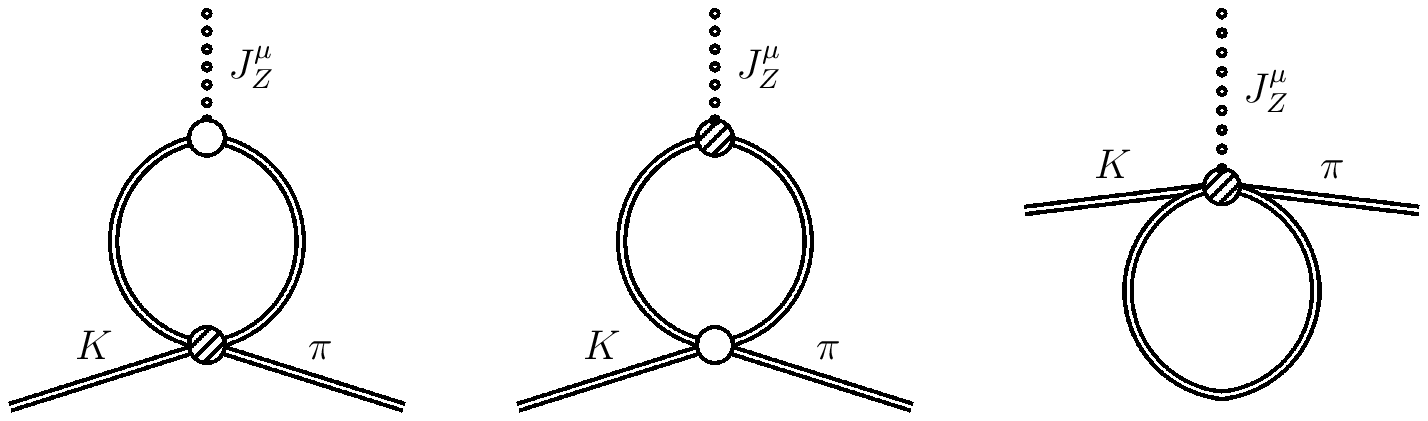} \vskip 0.3 cm 
b)\hskip 1 cm  \includegraphics[width=10 cm]{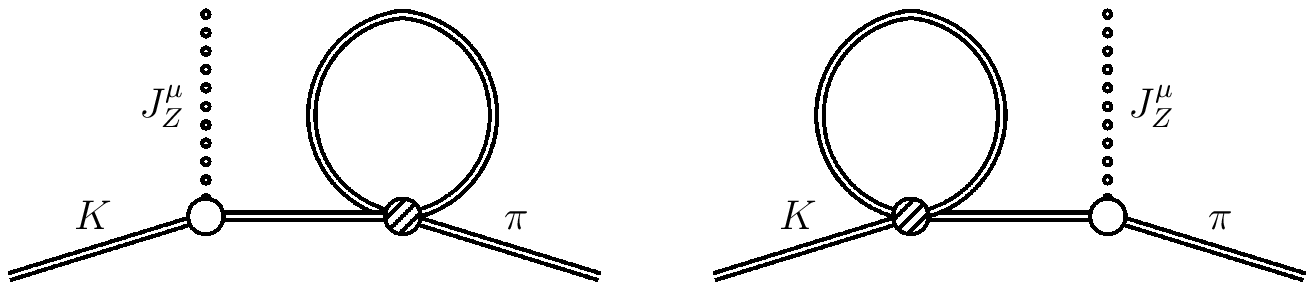} \vskip 0.3 cm 
c)\hskip 1 cm  \includegraphics[width=10 cm]{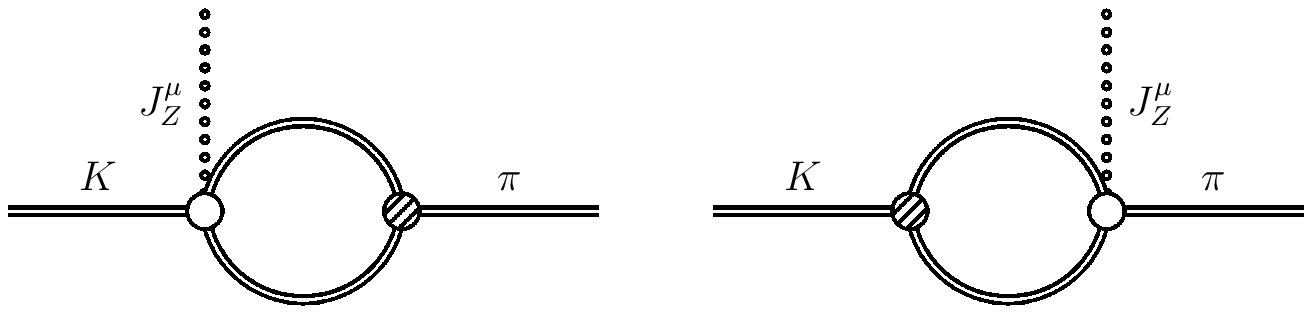} \vskip 0.3 cm 
d)\hskip 1 cm  \includegraphics[width=10 cm]{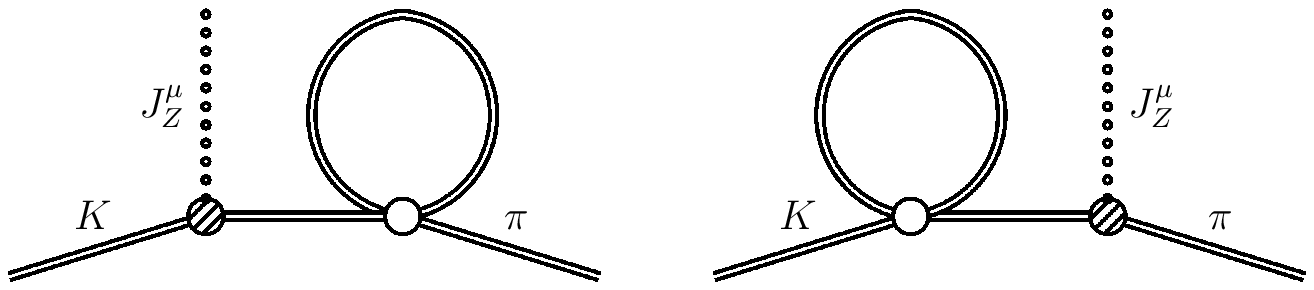} 
\end{center}
\caption{\it One-loop contributions to the $K\to \pi Z$ amplitudes within CHPT
in the pseudoscalar basis of Ref.~\cite{EPRgg}. Dashed and empty circles correspond 
to flavour-changing and flavour-conserving vertices, respectively. 
The wave-function renormalization diagrams (d) contribute only 
to $\cM_L^{+}$ (which is non-zero already at lowest order). }
\label{fig:oneloop}
\end{figure}

The $O(p^4)$ calculation of the $K \to \pi  Z$ amplitude involves several 
one-loop diagrams. However, a substantial simplification is obtained by 
performing the calculation in the basis of Ref.~\cite{EPRgg}, where 
the weak $O(p^2)$ mixing among pseudoscalar mesons is diagonalized,
and neglecting the $O(q^\mu)$ pole diagrams due to the 
$Z^{\mu}\partial_{\mu}\phi$ coupling (as at the tree-level, 
this completely removes the unknownw $a_1$ parameter from the relevant vertices). 
With these simplifications, the relevant one-loop diagrams 
are shown in figure~\ref{fig:oneloop}. 

In the case of the $\cM^+_{L}$ form factor,
the complete one-loop contribution can be decomposed as
\be
\left. \cM_{L}^{+(4)}\right|_{\rm loop}  = \frac{2}{(4\pi F)^2}\Big[
 \cT_{\rm a+b+c}^{0} 
 - (m_{K}^{2}-m_{\pi}^{2}) \cT_{\rm a+b+c}^{1} \Big]
 + \frac{1}{2}\left( \delta Z_{K}+\delta Z_{\pi}\right) \cM_L^{+(2)} 
\ee
where the subscripts refer to the labels of the 
diagrams in figure \ref{fig:oneloop}.
The explicit computation of the various terms yields
\bea
\cT_{\rm a+b+c}^{0} &=& 
A_{0}\left(  m_{K}^{2}\right)  +A_{0}\left(
m_{\eta_{8}}^{2}\right)  -\frac{1}{6}\left(  4m_{K}^{2}-q^{2}\right)
B_{0}\left(  q^{2};m_{K}^{2},m_{\pi}^{2}\right) \nn \\
&& +\frac{1}{12}\left(  4m_{K}^{2}-q^{2}\right)  B_{0}\left(  q^{2};m_{K}^{2},m_{K}^{2}\right)  
+\frac{1}{12}\left(  4m_{\pi}^{2}-q^{2}\right)
B_{0}\left(  q^{2};m_{\pi}^{2},m_{\pi}^{2}\right) 
\eea
\bea
\cT_{\rm a+b+c}^{1}  &=& 
\frac{1}{6}\left(  2+\frac{m_{K}^{2}}{m_{\pi}^{2}}\right)
B_{0}\left(  m_{\pi}^{2};m_{K}^{2},m_{\eta_{8}}^{2}\right)  +\frac{1}{6}
\left(  1-\frac{m_{K}^{2}}{m_{\pi}^{2}}\right)  B_{0}\left(  0;m_{K}^{2},m_{\eta_{8}}^{2}\right) \nn \\
&&  +\left(  \frac{2}{3}-\frac{m_{K}^{2}}{2m_{\pi}^{2}}+\frac{m_{K}^{2}-m_{\pi
}^{2}}{6q^{2}}\right)  B_{0}\left(  0;m_{K}^{2},m_{\pi}^{2}\right)  \nn\\
&& -\left(\frac{1}{3}+\frac{m_{K}^{2}-m_{\pi}^{2}}{6q^{2}}\right)  B_{0}\left(
q^{2};m_{K}^{2},m_{\pi}^{2}\right) 
+\frac{m_{K}^{2}}{2m_{\pi}^{2}}B_{0}\left(  m_{\pi}^{2};m_{K}^{2},m_{\pi}^{2}\right)
\eea
\bea
\delta Z_{\pi}  &=& - \frac{1}{(4\pi F)^2}\left[ \frac{2}{3} A_{0}\left(  m_{\pi}^{2}\right) 
  +\frac{1}{3}A_{0}\left(  m_{K}^{2}\right) \right] \\
\delta Z_{K}  &=&   - \frac{1}{(4\pi F)^2}\left[
 \frac{1} {4}A_{0}\left(  m_{\pi}^{2}\right) + \frac{1}{2}A_{0}\left(  m_{K}^{2}\right)
 +\frac{1}{4}A_{0}\left(  m_{\eta_{8}}^{2}\right)  \right]
\eea
where the expressions for the loop functions $A$ and $B$,
evaluated in dimensional regularization, are given in the Appendix. 

Interestingly, the complete expression of $\cM_{L}^{+(4)}|_{\rm loop}$ 
is finite and vanishes exactly in the $SU(3)$ limit 
($m_{K}=m_{\eta_{8}}=m_{\pi}$). The amplitude has a mild $q^2$ dependence 
and for $q^2> 4 m_\pi^2$ develops a small imaginary part
--related to the $K\to 3\pi$ intermediate state--
via the loop function  $B_{0}(q^{2};m_{\pi}^{2},m_{\pi}^{2})$. 
From the numerical point of view, the various terms 
in $\cM_{L}^{+(4)}|_{\rm loop}$ are separately of $O(1)$; 
however, there is a strong cancellation among them:
using the Gell-Mann Okubo (GMO) relation among pseudoscalar masses, 
we find 
\be
\left| \cM^{+(4)}_{L} \right|_{\rm loop}  < 0.01
\label{eq:cancell}
\ee
in all the allowed $q^2$ range. This cancellation is 
very sensitive to possible violations of the GMO relation,
but even for physical masses we find  a result 
substantially smaller than the tree-level value in Eq.~(\ref{eq:tree}).

\medskip

In the case of the $\cM^+_{V}$ form factor,
the result of the one-loop calculation yields
\bea
&& \left. \cM_{V}^{+(4)} \right|_{\rm loop} 
= \frac{2}{(4\pi F)^2} \left(1-\frac{4}{3}\sin^{2}\theta_{W}\right) \left[
 \frac{ m_{K}^{2}}{q^2} B_{0}\left(  0;m_{K}^{2},m_{K}^{2}\right)  
 +\frac{ m_{\pi}^{2}}{q^2} B_{0}\left(  0;m_{\pi}^{2},m_{\pi}^{2}\right) \right. \nn\\
&&\qquad  \left. +\frac{1}{4} B_{0}\left(  q^{2};m_{K}^{2},m_{K}^{2}\right)  
  \left(1-\frac {4m_{K}^{2}}{q^2} \right) 
+\frac{1}{4}B_{0}\left(  q^{2};m_{\pi}^{2},m_{\pi}^{2}\right)  \left( 1-\frac{4m_{\pi}^{2}}{q^2} \right) 
 +\frac{1}{3} \right] \qquad  \label{eq:MV4p}
\eea
As expected,  this result coincides (up to the overall normalization)
with the one-loop expression of the $K^{+}\rightarrow\pi^{+}\gamma$ 
amplitude obtained by Ecker, Pich and de Rafael \cite{EPR}. 
To make more explicit the connection with their result,  
it is sufficient to note that 
\be
\left. \cM_{V}^{+(4)}\right|_{\rm loop} 
  = \frac{1}{(4\pi F)^2} \left(1-\frac{4}{3}\sin^{2}\theta_{W}\right) 
\left[ D_{\varepsilon} - \log\frac{m_K m_\pi }{\mu^{2}} 
+ 3 \phi_\pi \left(q^2\right) + 3 \phi_K \left(q^2\right)    \right] \label{eq:MV4p_2} 
\ee
where $D_{\varepsilon}$ and $\phi_i(q^2)$  
--defined as in Ref.~\cite{EPR}-- are reported in the Appendix.
Contrary to the one-loop expression of  $\cM^+_{L}$,
the result in (\ref{eq:MV4p}) is not finite and 
does not vanish in the $SU(3)$ limit.

\medskip

Although not strictly necessary for practical purposes, 
we report here also the one-loop results for the neutral 
form factors, which are useful to investigate the 
$SU(3)$ properties of $K \to \pi  Z$ amplitudes.  
Due to the absence of tree-level contributions, 
the calculation of the neutral form factors is
somewhat simpler. We find 
\bea
&& \left. \cM_{L}^{0(4)}\right|_{\rm loop}  = - \frac{\sqrt{2}}{(4\pi F)^2}
\left\{ -\frac{5}{3}\left[  A_{0}\left(  m_{K}^{2}\right)-A_{0}\left(  m_{\pi}^{2}\right)  \right]
\right. \nn\\  
&&\qquad\qquad  +\frac{1}{6}\left(  4m_{K}^{2}-q^{2}\right)  \left[  B_{0}\left(  q^{2};m_{K}^{2},m_{K}^{2}\right)
-B_{0}\left(  q^{2};m_{K}^{2},m_{\pi}^{2}\right)  \right] \nn \\
&&\qquad\qquad +\left(  m_{K}^{2}-m_{\pi}^{2}\right)  \left[  \frac{1}{3}-2B_{0}\left(
m_{K}^{2};m_{\pi}^{2},m_{\pi}^{2}\right)  
 +  \frac{m_{K}^{2}-m_{\pi}^{2}+2q^{2}}{6q^{2}}B_{0}\left(  q^{2};m_{K}^{2},m_{\pi}^{2}\right) 
\right. \nn \\ 
&&\qquad\qquad \left.\left.-\frac{m_{K}^{2}-m_{\pi}^{2}+q^{2}}{6q^{2}}B_{0}\left(  0;m_{K}^{2},m_{\pi
}^{2}\right)  \right] \right\} \label{eq:ML0}
\eea
and 
\bea
 \left. \cM_{V}^{0(4)} \right|_{\rm loop} &=&  
 -\frac{1}{\sqrt{2} (4\pi F)^2} \left(1-\frac{4}{3}\sin^{2}\theta_{W}\right) 
\left[ D_{\varepsilon} -
\log\frac{m^2_K}{\mu^{2}} + 6 \phi_K \left(q^2\right)    \right]
 \label{eq:MV40_2} 
\eea
Also in this case the $\cM_{L}$ term vanishes in the $SU(3)$ limit and 
there is perfect analogy with the result of Ref.~\cite{EPR} for the vector
form factor. 
However, contrary to the charged case, Eq.~(\ref{eq:ML0}) is not finite 
beyond the $SU(3)$ limit. In particular, we find 
\be
\left. \cM_{L}^{0(4)}\right|_{\rm loop}^{\rm div}
 =  \frac{7 (  m_{K}^{2}-m_{\pi}^{2})}{\sqrt{2} (4\pi F)^2 }  D_\varepsilon
\ee
Concerning the finite parts, the one-loop expression of $\cM_{L}$ is $O(1)$,
it is almost constant, and it has a large absorptive part 
associated to the $K^0\to\pi^+\pi^-$ intermediate state.

\medskip

Combining all the one-loop results, the $SU(3)$ limit of the two 
amplitudes satisfy the relation 
\be
\left. \cA( K^+ \to \pi^+ \nu\bar\nu)^{\rm loop}_{Z} \right|_{SU(3)}
= -\sqrt{2} \left. \cA( K^0 \to \pi^0 \nu\bar\nu)^{\rm loop}_{Z} \right|_{SU(3)}
\equiv  \cA^{\rm loop}_{Z}~,
\ee
in perfect analogy with the $K^{+}\rightarrow\pi^{+}\gamma$ case.
In this limit, the (UV) scale dependence of the amplitude is given by 
\be
\mu^2 \frac{d}{d\mu^2} \cA^{\rm loop}_{Z}
= \left(1-\frac{4}{3}\sin^{2}\theta_{W}\right) \frac{G_F G_8}{16 \sqrt{2} \pi^2 } \times 
q^2 \left[ p^\mu \nu_l \gamma_\mu (1-\gamma_5) \nu_l \right]~.
\label{eq:scale_CHPT}
\ee

\subsection{Matching and $O(p^4)$ counterterms for $K \to \pi  Z$ amplitudes}
\label{sect:CT}

One of the key features of our result, is the fact that the scale dependence 
of the one-loop CHPT amplitude turns out to be proportional 
to $(1-\frac{4}{3}\sin^{2}\theta_{W})$. This fact is particularly 
welcome since it signals a short-distance behavior in agreement 
with the one derived at the partonic level. Indeed, the same coupling appears 
in the Wilson coefficient of $Q_{1l}^{(8)}$, namely the four-fermion 
dimension-8 operator corresponding to the $Z$-penguin contraction. 
Thus the short-distance behavior of the one-loop CHPT amplitude 
is perfectly compatible with the IR structure derived by the 
partonic calculation. 
To be more explicit, the UV scale dependence in (\ref{eq:scale_CHPT})
should be compared with
\be
\mu_{IR}^2 \frac{d}{d\mu_{IR}^2} \cA^{(8)}_Z = -
 \left(1-\frac{4}{3}\sin^{2}\theta_{W}\right) \frac{G^2_F \lambda 
[3C_1(\mu_c)+C_2(\mu_c)] }{12 \pi^2 } \times  q^2 \left[ p^\mu \nu_l \gamma_\mu (1-\gamma_5) \nu_l \right],
\label{eq:short_muZ}
\ee  
which follows from Eq.~(\ref{eq:A8summ}). As expected, the two expressions 
do not match exactly, given the non-perturbative QCD effects encoded 
in $G_8$, but they have the same kinematical dependence,
and the same parametrical dependence from the electroweak couplings. The 
second feature is a direct consequence of the matching condition 
which has been imposed on the $O(G_F^2p^2)$ Lagrangian in (\ref{eq:LW2_tot}).

As a consequence of the non-perturbative enhancement of $G_8$ 
($G_8 \gg G_F \lambda $), which is a manifestation of the $\Delta I=1/2$ rule, 
the scale dependence derived within CHPT is substantially larger 
with respect to the one obtained by the dimension-8 partonic Hamiltonian. 
This fact signals that, at least in the case of $Z$-mediated amplitudes,
the genuine long-distance contributions evaluated within CHPT represent 
the dominant effect.

\medskip

Within a pure CHPT approach, the scale dependence of the $\cL^{(2)} \times \cL^{(2)}$ 
one-loop amplitude must be exactly compensated  by that of the $O(p^4)$ local counterterms,  
whose renormalized finite parts encode possible short-distance contributions.  
In the present case, we can distinguish three types of $O(p^4)$ counterterms: 
\begin{enumerate}
\item[i.] the $L_i(\mu)$ of the strong Lagrangian \cite{GL}; 
\item[ii.] the $N_i(\mu)$ 
of the $O(G_F p^4)$ non-leptonic weak Lagrangian \cite{EKW}; 
\item[iii.] new  $O(G_F^2 p^4)$ local terms with explicit $l_\mu$ fields,
namely the  $O(p^4)$ generalization of the $O(p^2)$ terms in Eq.~(\ref{eq:newops}). 
\end{enumerate}
Unfortunately, only in the first two cases the finite parts of the 
counterterms are experimentally known. Considering only the 
contribution of these known couplings, we can write 
\bea
\left. \cM_{V}^{+(4)} \right|_{\rm CT} &=& 
\frac{4}{F^2} \left(1-\frac{4}{3}\sin^{2}\theta_{W}\right) 
\left[N_{14}(\mu)-N_{15}(\mu)+3L_9(\mu)\right] + O(N_i^\prime)  \qquad \nn \\
&\equiv& \frac{1}{(4\pi F)^2} \left(1-\frac{4}{3}\sin^{2}\theta_{W}\right) 
\left[ 1 -3 \frac{G_F}{|G_8|} a_+  +  \log\frac{m_K m_\pi}{\mu^{2}} \right] 
+ O(N_i^\prime)  \qquad  \label{eq:MV4p_CT} \\
\left. \cM_{V}^{0(4)} \right|_{\rm CT} &=& 
-\frac{\sqrt{2} }{ F^2}  \left(1-\frac{4}{3}\sin^{2}\theta_{W}\right) 
\left[ 2N_{14}(\mu)+N_{15}(\mu) \right] + O(N_i^\prime)  \qquad \nn \\
&\equiv& - \frac{1}{\sqrt{2} (4\pi F)^2 } \left(1-\frac{4}{3}\sin^{2}\theta_{W}\right) 
\left[1+  3 \frac{G_F}{|G_8|} a_S  + \log\frac{m^2_K}{\mu^{2}} \right] + O(N_i^\prime)  \qquad 
 \label{eq:MV40_CT} 
\eea
where $O(N_i^\prime)$ denote $\sin^{2}\theta_{W}$-independent 
and $\mu^2$-independent combination of counterterms which includes 
the new unknown couplings, as well as the unknown $a_1$ parameter
from the $O(p^4)$ strong and non-leptonic weak Lagrangian (i.~and ii.).  
According to the experimental data on $K^+\to \pi^+\ell^+\ell^-$ 
decays, the numerical value of the (finite) low-energy constant 
appearing in (\ref{eq:MV4p_CT}) is $a_+ \approx -0.6$ \cite{DEIP,BDI}.

\medskip

Concerning $\cM_L$, the counterterm structure is more complicated.
In the charged mode, where the tree-level result is different from zero, 
we expect terms correcting $G_8$ and the meson decay constants. 
From the konwn size of the latter, we naturally expect 
$O(25\%)$ corrections to the tree-level results, though we cannot 
exclude that strong cancellations similar to those leading 
to Eq.~(\ref{eq:cancell}) do occur.
It should also be noted that spurious $O(G_F^2 p^4)$ effects
from dimension-6 operators are still present in the 
$q^2$-independent part of both $\cM^{+(4)}_L$ and $\cM^{0(4)}_L$;
however, disentangling these effects from the genuine $O(G_F^2 p^4)$ 
long-distance corrections is beyond the scope of this work. 
Given these uncertainties on the counterterm structure of the 
 $\cM_L$ terms, in the numerical 
analysis of section~\ref{sect:Num} we shall assign a conservative  
$50\%$ error to the tree-level result for  $\cM^+_L$.

\subsection{$W$--$W$ amplitudes}

\begin{figure}[t]
\begin{center}
\includegraphics[width=12cm]{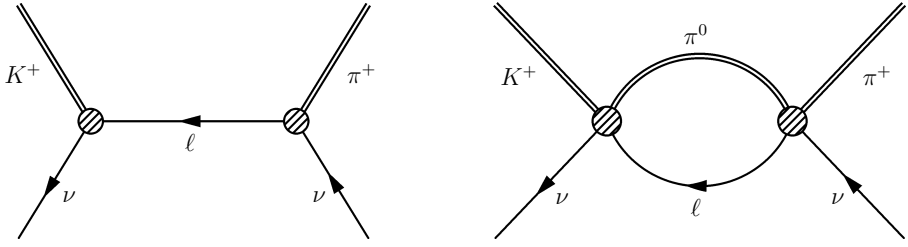}
\end{center}  
\caption{\it Tree-level and one-loop contributions to $K^+\to \pi^+\nu\bar\nu$ 
amplitudes induced by $W$--$W$ exchange, within CHPT.
Contrary to the case of figure~\ref{fig:fullmc}, here the 
index $\ell$ runs only over the first two families of leptons.}
\label{fig:WW}
\end{figure}

As can be seen from figure~\ref{fig:WW}, within CHPT 
the structure of FCNC amplitudes induced by two charged currents 
is substantially simpler than the $Z$-mediated case. On general grounds,
since the heavy $\tau$ lepton has already been integrated out 
at the level of the partonic Hamiltonian, we can decompose these 
contributions to the  $K^+ \to \pi^+\nu\bar\nu$ amplitude 
as follows
\be
\cA(K^+ \to \pi^+\nu\bar\nu)_{WW} =   G^2_F  F^2 \lambda
\sum_{l = e,\mu} \cM^{l}_{WW}~p^\mu \bar{\nu}_l \gamma_\mu (1-\gamma_5) \nu_l
\ee
The $O(p^2)$ tree-level contribution has already been analysed 
in Ref.~\cite{HL}; however, we disagree with their 
final result by a factor of 4. In particular, we obtain 
\be
\cM^{l(2)}_{WW} = 2 \left[ 1 + \frac{m_l^2}{t -m_l^2 } \right]  \approx  2 
\qquad\qquad (l=e,\mu) \label{eq:TB2}
\ee 
where $t=(p-p_\nu)^2$. Given the strong kinematical suppression of terms asymmetric in the two neutrino
momenta \cite{BI}, the approximate result in the r.h.s. of Eq.~(\ref{eq:TB2}) 
turns out to be an excellent approximation in the evaluation of the total decay rate.

The $O(p^4)$ one-loop calculation yields
\bea
\left. \cM^{l(4)}_{WW} \right|_{\rm loop} &=& \frac{1}{(4\pi F)^2} 
\left\{ (m_\pi^2-m_l^2) \left(2 -\frac{m_l^2}{2 t}\right) B_0(0; m_l^2, m_\pi^2) 
+4 A_0(m_l^2) -A_0(m_\pi^2) \right. \nn\\
&& \left. +\left[4(m_\pi^2-t) + (m_l^2-m_\pi^2+t)
\left(2 -\frac{m_l^2}{2 t}\right)\right]B_0(t; m_l^2, m_\pi^2) \right\} \qquad 
\eea
and in the limit $m_l \to 0$ becomes (see Appendix):
\bea
\left. \cM^{l(4)}_{WW} \right|_{\rm loop} & \stackrel{m_l \to 0}{\longrightarrow} &
\frac{1}{(4\pi F)^2} \left[ (3m_\pi^2 -2 t) 
\left( D_{\varepsilon} -\log\frac{m_\pi^2}{\mu^2} \right) \right. \nn \\
&& \left. +5 m_\pi^2 -4 t + 2 \frac{(m_\pi^2-t)^2}{t}\log\left(1-\frac{t}{m_\pi^2}\right)
\right] \label{eq:box_LD}
\eea
As expected, the one-loop result is divergent. Within a pure CHPT
calculation, such divergence is cured by appropriate $O(G_F^2 p^4)$ 
counterterms. Unfortunately, we cannot determine the finite part of these
counterterms, neither from first principles nor from data. However, 
a reasonable estimate of their size can be obtained by imposing the 
matching of the scale dependence with the short-distance partonic 
calculation. Since in this case there are no sizable non-perturbative effects 
associated to the $\Delta I=1/2$ rule (as in the $Z$-mediated case), 
we expect a good numerical matching.
From Eq.~(\ref{eq:box_LD}) it follows
\bea
\mu^2 \frac{d}{d\mu^2} \cA^{\rm loop}_{WW} &=& \frac{G_F^2 \lambda}{16 \pi^2} (3m_\pi^2 -2t) 
\left[ p^\mu \nu_l \gamma_\mu (1-\gamma_5) \nu_l \right]~ \nn\\
 &=& - \frac{G_F^2 \lambda}{8 \pi^2 } \left[ p\cdot k + O(m_\pi^2)
 \right] \left[ p^\mu \nu_l \gamma_\mu (1-\gamma_5) \nu_l \right] ~+~ 
O\left[ p\cdot (p_\nu-p_{\bar\nu} ) \right]~,\qquad 
\eea
while from Eq.~(\ref{eq:A8summ}) we get 
\be
\mu_{IR}^2 \frac{d}{d\mu_{IR}^2} \cA^{(8)}_{WW} =  \frac{G^2_F\lambda }{2\pi^2}
\left[ {\hat B}_{2} (p\cdot k) + O(m_q) \right]
\left[ p^\mu \nu_l \gamma_\mu (1-\gamma_5) \nu_l \right]~.
\label{eq:short_WW}
\ee  
As can be noted, the two expression have the same kinematical structure 
(within the approximations employed) and the scale dependence can be 
matched with an appropriate choice of the hadronic parameter
defined in Eq.~(\ref{eq:hatB2}), namely  ${\hat B}_{2} \approx 1/4$.
This result allows us to have a full control on the total 
$W$--$W$ amplitude, with the inclusion of the contribution 
of the dimension-8 partonic operator. Neglecting $O(p^4)$ terms not 
enhanced by the large factor $\log (m_c^2/m_\pi^2)$, we finally 
obtain
\be
\cM^{(4)}_{WW}~ = ~  2 - \frac{1}{16\pi^2 F^2} (m_K^2-q^2) \log \frac{m_c^2}{m_\pi^2}
+ O\left(\frac{m^2_K}{16\pi^2 F^2}\right) \qquad\qquad (l=e,\mu) 
\label{eq:WWtot}
\ee 
Note that, despite the large-log enhancement, 
the $O(p^4)$ term turns out to be smaller than 
the tree-level contribution. It is also worth to stress that  
the overall term in (\ref{eq:WWtot}) is smaller that the estimate 
of the dimension-8 contribution only presented in Ref.~\cite{Falk},
which was based on na\"ive dimensional analysis.

\section{Numerical analysis for  $\cB(K^+\to\pi^+\nu\bar\nu)$}
\label{sect:Num}

Collecting the results of the previous section, 
we can finally estimate the {\em complete} 
subleading $Z$-- and  $W$--$W$--mediated contributions 
to the $K^+\to\pi^+\nu\bar\nu$ amplitude
(including both dimension-8 and long-distance effects).  
We express them through the coefficients $P_Z(q^2)$ and 
$P_{WW}(q^2)$, defined by 
\be
\cA( K^+ \to \pi^+ \nu \bar\nu ) =
- \frac{G_F}{{\sqrt 2}}\frac{\alpha~\lambda^5 }{ 2\pi \sin^2\theta_W}
\sum_{l=e,\mu,\tau} \left[ P_Z(q^2) + P^l_{WW}(q^2)\right]
 (p+k)^\mu \bar{\nu}_l \gamma_\mu (1-\gamma_5) \nu_l 
\ee
The normalization of $P_Z(q^2)$ and $P_{WW}(q^2)$ is such that they can  
easily be compared with the leading dimension-6 charm contribution
in Eq.~(\ref{eq:Pc_d}). In particular, the combination
\be
\delta P_{c,u} = \frac{1}{3} \sum_{l=e,\mu,\tau} \left\langle  P_Z(q^2) + P^l_{WW}(q^2)\right\rangle
\ee
where $\langle\rangle$ denotes an appropriate average over the phase space, 
represents the correction to be added to the leading coefficient 
\be
 P^{(6)}_c = \frac{1}{\lambda^4} 
\left[\frac{2}{3}X^e_c +\frac{1}{3}X^\tau_{c}\right] = 0.39 \pm 0.07~,
\label{eq:P06}
\ee
whose numerical value corresponds to the present NLO accuracy
of ${\cal H}^{(6)}_{eff}$ \cite{BB_new,Buras:2004}.

According  to the results in Eq.~(\ref{eq:tree}), (\ref{eq:MV4p_2}) and 
(\ref{eq:MV4p_CT}), we find 
\be 
P_Z(q^2) \, = \,  - \frac{ \pi^2 F^2  {\rm sgn}(G_8) }{\sqrt{2}\lambda^5  M_W^2} 
\left[ \frac{4 |G_8|}{G_F} -  \frac{3 a_+ q^2}{16 \pi^2 F^2}  \left(1 - \frac{4}{3} \sin^{2}\theta_{W}\right)
+ O(N_i^\prime, q^4) \right]
\ee
while for the $W$--$W$ contribution we get
\bea 
P^{e,\mu}_{WW}(q^2)  & = &   - \frac{ \pi^2 F^2   }{\lambda^4  M_W^2}
\left[  2 - \frac{1}{16\pi^2 F^2} (m_K^2-q^2) \log \frac{m_c^2}{m_\pi^2} 
+ O\left(\frac{m^2_K}{16\pi^2 F^2}\right) 
 \right]  \no\\
P^{\tau}_{WW}(q^2)  & = & - \frac{(m_K^2-q^2)}{32\lambda^4  M_W^2}  f\left( m_c^2/m_{\tau}^2\right) 
\eea
In the case
of $P_Z(q^2)$, we have explicitly pointed out the dependence from ${\rm sgn}(G_8)$, 
which has to be estimated starting from the partonic four-quark Hamiltonian. 
As discussed in Ref.~\cite{BDI}, employing the factorization approximation
leads to $G_8 < 0$. From the numerical point of view, the sum of the 
$O(p^4)$ terms encoded in $P_Z(q^2)$ and $P_{WW}(q^2)$ turn out to be 
about $20\%$ of the  $O(p^2)$ terms, in good agreement with na\"ive
chiral counting. However, as discussed at the end of section~\ref{sect:CT},
we have not been able to estimate all the possible $O(p^4)$ contributions.
For this reason, we believe that the most conservative approach for the 
numerical analysis is obtained by fixing the central value of $\delta P_{c,u}$
from the complete $O(p^2)$ result, and attribute to it a $50\%$ error:
\be 
\delta P_{c,u} \, \approx  \,  \frac{ \pi^2 F^2   }{\lambda^4  M_W^2}
\left[ \frac{ 4 |G_8|}{\sqrt{2}\lambda  G_F} - \frac{4}{3} \right] \,   = \, 0.04 \pm 0.02 
\label{eq:PC}
\ee  
In summary, the subleading contributions of $O(G^2_F \Lambda_{\rm QCD}^2)$
to the $K^+ \to \pi\nu\bar\nu$ amplitude turns out to 
be a {\em constructive} $10\%$ correction  
with respect to the leading charm contribution
\be
 P^{(6)}_c \to  P^{(6)}_c +\delta P_{c,u} \qquad \qquad \delta P_{c,u} = 0.04 \pm 0.02
\label{eq:PC2}
\ee
This correction implies a $\approx 6 \%$ increase of the SM 
prediction of  $\cB(K^+\to\pi^+\nu\bar\nu)$.  

The overall effect of the $O(G^2_F \Lambda_{\rm QCD}^2)$ corrections turns 
out to be smaller than the error of the dimension-6 term at the 
NLO level, thus it was justified to neglect these subleading terms 
at this level of accuracy (as for instance done in the recent analysis 
of Ref.~\cite{Buras:2004,DI}). However, these subleading terms are 
not negligible in view of a NNLO analysis 
of the dimension-6 contribution. 

\section{Conclusions}
\label{sect:concl}
   
We have presented a comprehensive analysis of the $O(G_F^2  \Lam^2)$
contributions to $K \to \pi \nu\bar\nu$ amplitudes not described by the leading 
dimension-6 effective Hamiltonian. These include both the effects 
of dimension-8 four-fermion operators generated at the charm 
scale, and the genuine long-distance contributions which can be described 
within the  framework of CHPT. As we have shown, these two type of 
effects are closely correlated.  The main results of our analysis 
can be summarized as follows:
\begin{itemize}
\item{} The dominant contributions are the  $O(G_F^2 p^2)$ tree-level
amplitudes which can be computed within CHPT. A consistent evaluation 
of these amplitudes requires the introduction of new chiral operators
of $O(G_F^2 p^2)$, in addition to those already present in the non-leptonic 
weak chiral Lagrangian. These operators, which are needed to cancel 
spurious tree-level FCNCs and to ensure a correct UV behavior of the
chiral amplitudes, have not been considered in the 
previous literature \cite{Lu,Geng,Fajfer}. 
\item{} The introduction of the new $O(G_F^2 p^2)$ operators 
has allowed us to obtain a consistent matching between the CHPT one-loop 
amplitudes of $O(G_F^2 p^4)$ and the contributions of the 
dimension-8 four-fermion Hamiltonian. Thanks to this matching, 
we have been able to estimate more precisely both these 
sources of subleading contributions. In particular, we have 
estimated the hadronic matrix element of all the 
dimension-8 partonic operators of Ref.~\cite{Falk}, 
strongly reducing this source of uncertainty. 
From the numerical point of view, both the $O(G_F^2 p^4)$ chiral
amplitudes and the dimension-8 partonic operators induce 
very small effects on $K \to \pi \nu\bar\nu$ amplitudes 
(corrections below the $1\%$ level).
\item{} In the $K^+\to\pi^+\nu\bar\nu$ case, 
the leading corrections due to the  $O(G_F^2 p^2)$
chiral amplitudes amount to about 10\% of the dimension-6 
charm contribution, or about 3\% of the total SM  amplitude. 
Their effect can be efficiently encoded in the 
standard analysis of $\cB(K^+\to\pi^+\nu\bar\nu)$
be means of Eq.~(\ref{eq:PC2}). The size of these effects 
can easily be understood by noting that they scale as
$(\pi F/m_c)^2 \approx 5\% $ with respect to the leading 
charm contribution and are partially enhanced by the 
$\Delta I=1/2$ rule (see Eq.~(\ref{eq:PC})). 
These subleading terms are not negligible 
in view of a NNLO analysis of the dimension-6 contribution. 
\end{itemize}

\section*{Acknowledgments}
We thank Andrzej Buras, Gerhard Ecker, Guido Martinelli,
Eduardo de Rafel, and Paolo Turchetti for useful comments and discussions. 
This work is partially supported by IHP-RTN, 
EC contract No.\ HPRN-CT-2002-00311 (EURIDICE).

\appendix
\section{Loop functions}
Following standard conventions, we define
\bea
B_{0}\left(  q^{2},m_{1}^{2},m_{2}^{2}\right)  &=& -i~(4\pi)^2 ~\mu^{\varepsilon} 
\int\frac{d^{d}k}{\left(  2\pi\right)^{d}}\frac{1}{\left(  k^{2}-m_{1}^{2}\right)  
\left(  \left(  k-p\right)^{2}-m_{2}^{2}\right)  } \\
A_{0}(m^{2})  &=&
-i  ~(4\pi)^2 ~\mu^{\varepsilon}\int\frac{d^{d}k}{\left(  2\pi\right)^{d}}\frac{1}{\left(  k^{2}-m^{2}\right) }
\eea
where $\varepsilon=4-d$. The following identities holds
\bea 
B_{0}\left(  0,m^{2},m^{2}\right)   
&=& \frac{A_{0}\left(  m^{2}\right)}{m^{2}}-1= 
D_{\varepsilon}-\log\frac{m^{2}}{\mu^{2}}  \nn \\
B_{0}\left(  0,m_{1}^{2},m_{2}^{2}\right)   &=&  
D_{\varepsilon}+1-\frac{1}{m_{1}^{2}-m_{2}^{2}}
\left(  m_{1}^{2}\log\frac{m_{1}^{2}}{\mu^{2}}-m_{2}^{2}\log\frac{m_{2}^{2}}{\mu^{2}}\right)  
\nn \\
&=& \frac{A_{0}\left(  m_{1}^{2}\right)  -A_{0}\left(  m_{2}^{2}\right)}{m_{1}^{2}-m_{2}^{2}} \nn \\
B_{0}\left(  q^{2},m^{2},m^{2}\right)  &=&  
D_{\varepsilon}-\log\frac{m^{2}}{\mu^{2}}
+H_{1}\left(  \frac{q^{2}}{m^{2}}\right) 
\eea
where $D_{\varepsilon}=2/\varepsilon-\gamma+\log4\pi$ and 
\be
H_{1}\left(  a\right)  =\left\{
\begin{array}
[c]{lll}%
2-2\sqrt{4/a-1}\arctan\dfrac{1}{\sqrt{4/a-1}} &  & a<4\\
2-\sqrt{1-4/a}\left(  \log\dfrac{1+\sqrt{1-4/a}}{1-\sqrt{1-4/a}}-i\pi\right)
&  & a>4
\end{array} 
\right.
\ee
The one-loop function defined in Ref.~\cite{EPR} is 
\be
\phi_i (q^2)  = -\frac{4}{3}\frac{m_i^{2}}{q^{2}} 
+\frac{5}{18}+\frac{1}{3}\left(  \frac{4m_i^{2}}{q^{2}}-1\right)  
\frac{2-H_{1}\left(  q^{2}/m_i^{2}\right)  }{2} 
\ee
and for small $q^2$ can be expanded as 
\be
\phi_i (q^2)  = -\frac{1}{6}\left[ 1+ \frac{q^2}{10m_i^2} +O\left(\frac{q^4}{m_i^4}\right)\right]
\ee


\end{document}